\documentclass[preprint,11pt]{aastex}

\newcommand{\etal}{{et al.}}

\newcommand{\Msun}{\>{\rm M_{\odot}}}
\newcommand{\Lsun}{\>{\rm L_{\odot}}}

\begin{document}
\title{Circumnuclear Shock And Starburst In NGC 6240 : Near-IR Imaging And Spectroscopy With Adaptive Optics}
\author{Tamara Bogdanovi\'c \altaffilmark{1}, Jian Ge \altaffilmark{2}}
\affil{Department of Astronomy \& Astrophysics, The Pennsylvania State University, 
University Park, PA 16802}
\author{Claire E. Max\altaffilmark{3}}
\affil{Institute of Geophysics and Planetary Physics, L-413 \\
Lawrence Livermore National Laboratory, Livermore, CA 94550 \\ and 
Center for Adaptive Optics, University of California, Santa Cruz,
CA 95064}
\author{Lynne M. Raschke\altaffilmark{4}}
\affil{Department of Astronomy and Astrophysics \\
and Center for Adaptive Optics, University of California, Santa Cruz, CA 95064}

\altaffiltext{1}{tamarab@astro.psu.edu}
\altaffiltext{2}{jian@astro.psu.edu}
\altaffiltext{3}{max1@llnl.gov}
\altaffiltext{4}{lynne@ucolick.org}

\begin{abstract}
  
 We have obtained adaptive optics, high spatial resolution (0.15
 arcsecond) K-band spectra and images of the region around the two
 active nuclei in NGC 6240 which show the presence of circumnuclear
 shocks. The data are consistent with the thermal excitation mechanism
 being the dominant one in the nuclear region. UV fluorescence and
 associative detachment may also contribute to the fraction of the
 energy emitted through molecular hydrogen transitions. The near-IR
 continuum emission appears closely associated with the two active
 nuclei. The morphological similarities between the near-IR images and
 the Chandra X-ray images indicate the same mechanisms may be
 responsible for the emission in near-IR and X-ray band.

\end{abstract}

\keywords{galaxies: individual (NGC 6240) --- galaxies: nuclei --- galaxies: starburst --- techniques: high angular resolution}

\section{Introduction}
\subsection{NGC 6240}

	NGC 6240 is one of the best studied composite galaxies,
	hosting a luminous starburst and two active galactic nuclei
	(AGNs) obscured by gas and dust. With a bolometric luminosity
	of $6\times10^{11}\Lsun$ \citep{Rieke}, emitted predominantly
	in the infrared, this galaxy is a borderline case of the
	ultra-luminous infrared galaxies (ULIRGs,
	$L_{ULIRG}>10^{12}\Lsun$). Moreover, NGC 6240 has the largest
	known near-infrared molecular hydrogen line luminosity,
	$L(H_{2})_{tot}\approx 10^{9}\Lsun$ \citep{Draine}. Many of
	the ULIRGs are objects with a large amount of molecular
	hydrogen gas. Their extraordinary infrared luminosity
	originates either in a powerful starburst or in a gas embedded
	AGN; NGC 6240 is an object with clear signatures of both.

	We observed the two distinct nuclei at high angular resolution
	with adaptive optics. Both, the northern and southern nucleus
	(hereafter the N nucleus and the S nucleus respectively) are
	surrounded by bright gas components. The S nucleus shows
	strong nuclear activity and has been suggested to be the
	source of the AGN hard X-ray component
	\citep{Iwasawa,Vignati,Ikebe}. Being embedded in the gas,
	dust, and powerful starburst, the AGN is not directly
	observable at energies lower than $10\,$keV. Recent Chandra
	ACIS data \citep{Komossa1} reveal two active nuclei in hard
	X-ray emission, coincident with the optical and IR nuclei of
	NGC 6240.

        NGC 6240 was observed in the near-infrared by many authors in
        the past,
        \citet{Wade,Rieke,Lester2,Elston,Herbst,vdWerf,Lester1,Lancon,Oliva,Ohyama,Scoville,Tecza},
        with the latest works setting the quality and resolution
        criteria for near-IR spectroscopy and imaging of this object.
        An important goal of the past 20 years of study of this galaxy
        has been to determine and explain the dominant emission
        mechanism. This objective has been difficult to meet because
        both very accurate line ratios and spatial information about
        the emission lines are needed to draw firm conclusions. In the
        next section we review some of the proposed excitation
        mechanisms.

\subsection{Proposed Excitation Mechanisms}

	There are four excitation mechanisms proposed by different
	authors to explain the dominant source of energy emitted in
	the near-infrared from the nuclear region of NGC 6240.

        {\it Thermal emission driven by collisional excitation}
        \citep{Sugai,Lester2,Ohyama}. This type of excitation arises
        from the collision of $H_{2}$ molecules with other hydrogen
        molecules, atoms, and energetic electrons. NGC 6240 is a merger
        galaxy and the presence of shocks is expected both as a
        consequence of tidal effects and starburst driven
        superwinds. 

        {\it Ultraviolet fluorescence} \citep{Tanaka}. This mechanism
        is expected to arise in galaxies hosting a young starburst
        population of supergiant stars. Starburst galaxies typically
        have a large amount of gas that responds to the UV radiation.
        The gas shows a characteristic disrupted morphology and strong
        infrared emission.

       {\it X-ray heating by the central source and supernovae}
       \citep{Draine}. NGC 6240 is known to have two hard X-ray nuclei
       embedded in gas \citep{Komossa1}. The central, high-temperature
       source ionizes the gas in its vicinity and creates high energy
       electrons. These further participate in collisions with $H$ and
       $H_{2}$. Since the column density of the gas in the nuclear
       region of NGC 6240 is high ($N_{H}\simeq 10^{21-22}\, {\rm
       cm^{-2}}$; Komossa et al. 2002), X-ray emission is expected to
       be mostly thermalized and re-emitted in the IR
       continuum. Alternatively X-rays can be produced by supernovae
       and high-velocity cloud-cloud collisions. The emission from
       this mechanism depends on the power of the source and column
       density of the gas.

       {\it A formation pumping mechanism, driven by dissociative
       shocks} \citep{Mouri}. Fast, dissociative shocks passing
       through the molecular gas leave behind $H$ atoms and ions that
       recombine into excited $H_{2}$ and give rise to formation
       pumping. This process is also called {\it associative
       detachment}.

       Our observations allow further investigation into which
       mechanisms contribute the dominant fraction of energy radiated
       in near-IR. In $\S 2$ we describe the observations and data
       reduction procedures applied to the spectra and images. In $\S
       3$ we present the results; in $\S 4$ we discuss their
       implications and give concluding remarks in $\S 5$.

\section{Observation and Data Reduction}
	
	In June 2000 and August 2001 we obtained high spatial
	resolution ($\sim0.15$ arcsecond) near-IR images and spectra
	of NGC 6240 with the Lick 3 meter telescope and the natural
	guide star adaptive optics system \citep{Gavel}. The
	observations were carried out with the PICNIC $256\times 256$
	IR array IRCAL camera \citep{Lloyd}. The plate scale was
	75.6$\pm$0.2 milliarcsec per pixel with a total field of view
	of 19.4 arcsec. For our spectroscopic observations we used a
	CaF$_{2}$-grism \citep{Ge}, which covered a wavelength band
	$1.9-2.4\,{\rm \mu m}$, with a wedge angle of 30.95 degrees.
	One pixel corresponded to approximately 19.5\AA\ along the
	dispersion direction. The spectral resolution achieved was
	about $750\,{\rm km\,s^{-1}}$, or
	R=$\lambda/\Delta\lambda$=400. The total on-source
	spectroscopic exposure time was 6900 seconds (23 exposures,
	300 seconds integration each) (Table~\ref{table1}) and 4465
	seconds total for imaging (Table~\ref{table2}).

\subsection{Spectra of the NGC 6240 Nuclear Region}

\begin{figure}
\epsscale{0.8}
\plotone{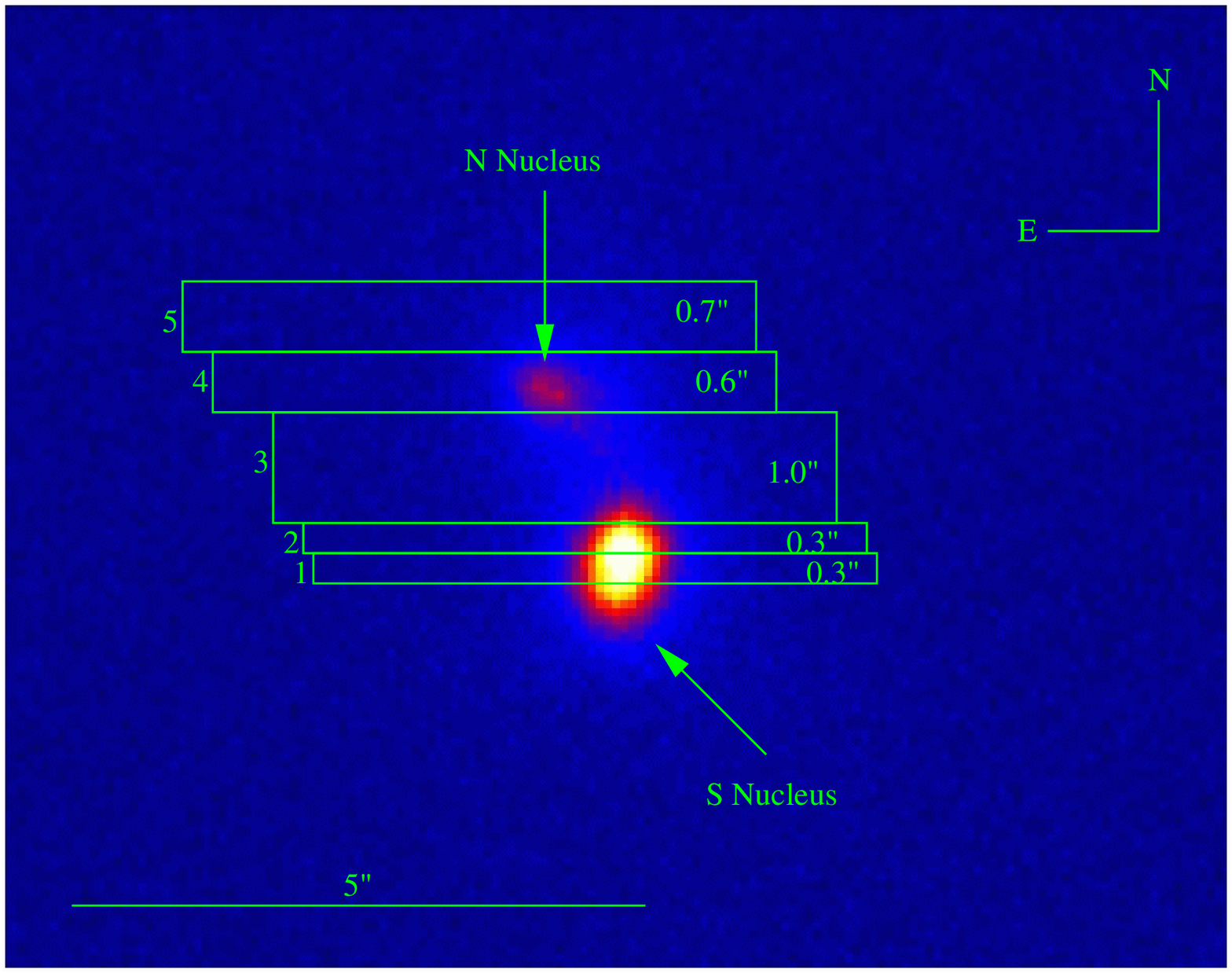}
\figcaption[bogdanovic.fig1.eps]{Ks-band image of the nuclear region with the scheme of five regions on the sky associated with 5 different spectral frames. The length of all boxes, marking the regions on the sky, is 5 arcseconds and the width is as marked on the figure.\label{Ks}}
\end{figure}

	The angular separation of the two galactic nuclei measured in
	the H and K bands is about 1.8 arcsec \citep{Lira}. The slit
	was initially positioned on the center of the S nucleus, the
	brighter one, and then progressively shifted towards the N
	nucleus. The instrumental slit width of 0.15 arcseconds and
	slit positioning allowed us to obtain coverage of the nuclei
	and the region between them. Spatially close frames, with
	similar spectral features were then combined in order to
	obtain higher signal to noise ratios for the five resulting
	spectral frames. The regions on the sky associated with every
	combined spectral frame are boxes of approximate sizes
	0.3$\times$5.0, 0.3$\times$5.0, 1.0$\times$5.0, 0.6$\times$5.0
	and 0.7$\times$5.0 arcseconds, positioned from the S towards N
	nucleus, respectively. The area of each box is defined by the
	slit widths (which depends on how many spectra are combined
	for particular frame) and slit length of 5 arcseconds. The
	scheme of positions of the boxes on the sky, associated with
	the five spectral frames, is shown in Figure~\ref{Ks}.

	Data reduction was performed using the IRAF software
	package. The sky frames, obtained right after the target
	exposure, were subtracted from all object frames. The data
	were flat fielded using dome flat lamp exposures, obtained at
	the end of each observation. For the spectral calibration we
	used a neon lamp, which has enough strong emission lines in
	the wavelength interval of interest to allow calibration. The
	flux units are arbitrary, as common for spectra obtained with
	adaptive optics system, where absolute flux values are
	uncertain. Later (Section 3.4) we base our analysis on
	relative line intensities. For this purpose we fit every line
	with a Gaussian and calculate its equivalent width. The
	equivalent width of a line sensitively depends on the adopted
	continuum level. We estimate the continuum level in the
	vicinity of each emission line. This helps us to minimize the
	error caused by the noise in the continuum not captured by the
	continuum fitting.
	
	The atmospheric absorption correction was carried out with the
	A0 V star HD 203856, observed on the same night as the object,
	by dividing the target spectra by the A0 V star spectrum. The
	Br$\gamma$ absorption region in the spectrum of the star was
	approximated with a Gaussian and then removed.

	We estimate the error for the calculated line ratios, from the
	error contribution for both lines. The uncertainty in the
	measurement of relative intensity of a single line is
	contributed by the finite signal to noise ratio and finite
	resolution.

\subsection{Images of NGC 6240 Nuclear Region}
 
	We obtained three sets of images: with the $Ks$ band filter,
	and $H_{2}$ and $Br\gamma$ narrow band filters, with exposure
	times of 60, 1722 and 2633 seconds, respectively
	(Table~\ref{table2}). The Ks filter has a central wavelength
	of $\lambda_{c}$=2.150$\,{\rm \mu m}$, and full width half
	maximum (FWHM) of 0.320$\,{\rm \mu m}$. The $H_{2}$ and
	$Br\gamma$ filters have central wavelengths of
	$\lambda_{c}$=2.125$\,{\rm \mu m}$ and
	$\lambda_{c}$=2.167$\,{\rm \mu m}$, respectively, and FWHM of
	0.020$\,{\rm \mu m}$. The latter two filters have
	approximately the same transmission, close to 80 per cent.

	Due to the redshift of the object (V$_{sys}$=7339$\,{\rm
	km\,s^{-1}}$; Downes, Solomon \& Radford 1993) the $H_{2}$
	filter samples the continuum emission and the Br$\gamma$
	filter samples the redshifted molecular hydrogen emission line
	$1-0S\,(1)$ as well as the underlying continuum. The $H_{2}$
	v=$2-1S(3)$ emission line, observed at a
	$\lambda_{obs}$=2.1243$\,{\rm \mu m}$ object frame, may
	contaminate the continuum emission observed with the $H_{2}$
	filter. However, the expected contamination effect should be
	small, since the line intensity of this transition in NGC 6240
	is very weak (Ohyama et al. 2000; see Section 3.4). The sky was
	subtracted from every object frame. The twilight flats taken
	at the beginning of the evening were used for the flat
	fielding of object frames.

  	We used Keck NIRSPEC and NIRC2 spectra, with spectral
  	resolution R=2000 and 0.4 arcsecond slit, taken by one of us
  	(C.Max) as an independent check of relative brightness scale
  	for images obtained with the two narrowband filters. The
  	$1-0S\,(1)$ emission line was resolved in each of two
  	spectra. Both spectra showed that the peak brightness of the S
  	nucleus is about a factor of 1.3 higher in the Br$\gamma$
  	filter bandpass (comprised of line plus continuum emission)
  	than in the $H_{2}$ filter bandpass (contributed by the
  	continuum emission). The scaling factor confidence is about
  	3$\sigma$. We rescaled the $H_{2}$ band image in order to make
  	its peak brightness at the Southern nucleus equal to 1/1.3
  	times that seen in Br$\gamma$, and then subtracted the
  	rescaled image from the Br$\gamma$ image. As a result we
  	obtained an image that emphasizes the molecular emission
  	region (Fig~\ref{narrow_band}a). The images show a region
  	about $11.34\times 9.07$ arcseconds ($\sim5.5\times 4.4\,$kpc)
  	in size. The images are overlaid with a linear isocontour in
  	order to highlight the morphology of the nuclear region. In
  	Fig~\ref{narrow_band}a we also mark the locations of the N and
  	S nuclei as seen in the near-IR continuum image in
  	Fig~\ref{narrow_band}b.

	Figure~\ref{xray_contour} shows an X-ray logarithmic
	isocontour image of binned counts obtained from the Chandra
	Data Archive. The Chandra observation was carried out with
	HRC-I microchannel plate detector \citep{Murray} in the
	$0.08-10.0\,$keV energy band with exposure time of 8.8
	ksec. The image size is $23.7\times 18.7$ arcseconds.
 	 
\section{Results}
\subsection{Ks Broad Band Image and Narrow Band Images}

	 Figure~\ref{Ks} shows the Ks-band image, obtained with the Ks
	 broad band filter, transparent to near-IR photons in a
	 wavelength range of 1.99$-$2.31$\,{\rm \mu m}$. The flux in
	 this wavelength range is contributed by the emission from the
	 continuum and molecular hydrogen lines. Our Ks-band image
	 closely resembles to the high resolution images reported by
	 \citet{Scoville}, taken in the 1.1$\,{\rm \mu m}$ and
	 2.2$\,{\rm \mu m}$ rest frame. Two compact nuclei are easily
	 distinguished on the image. Also noticeable is the faint,
	 extended emission around the nuclei.

\begin{figure}
\epsscale{1.1}
\plottwo{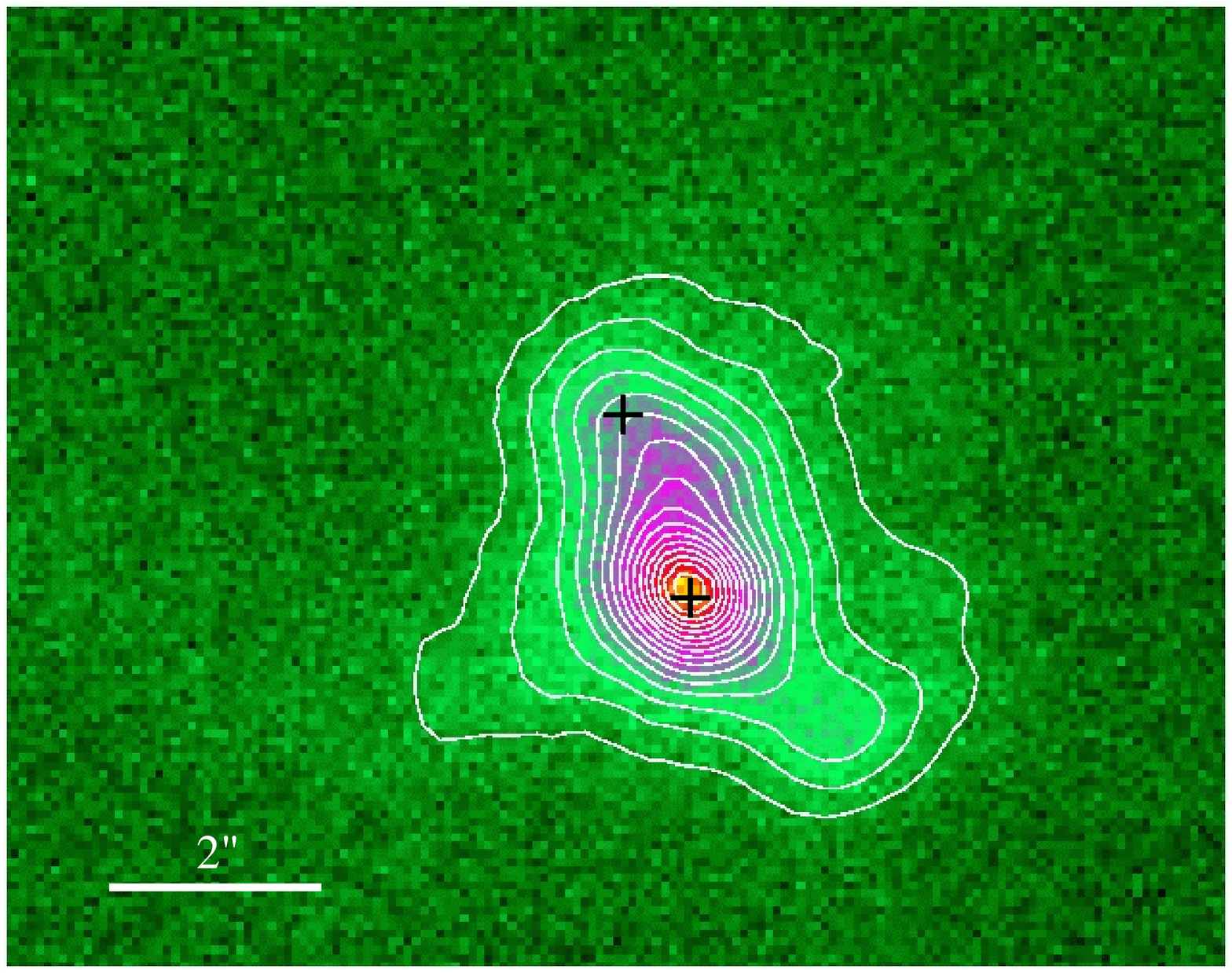}{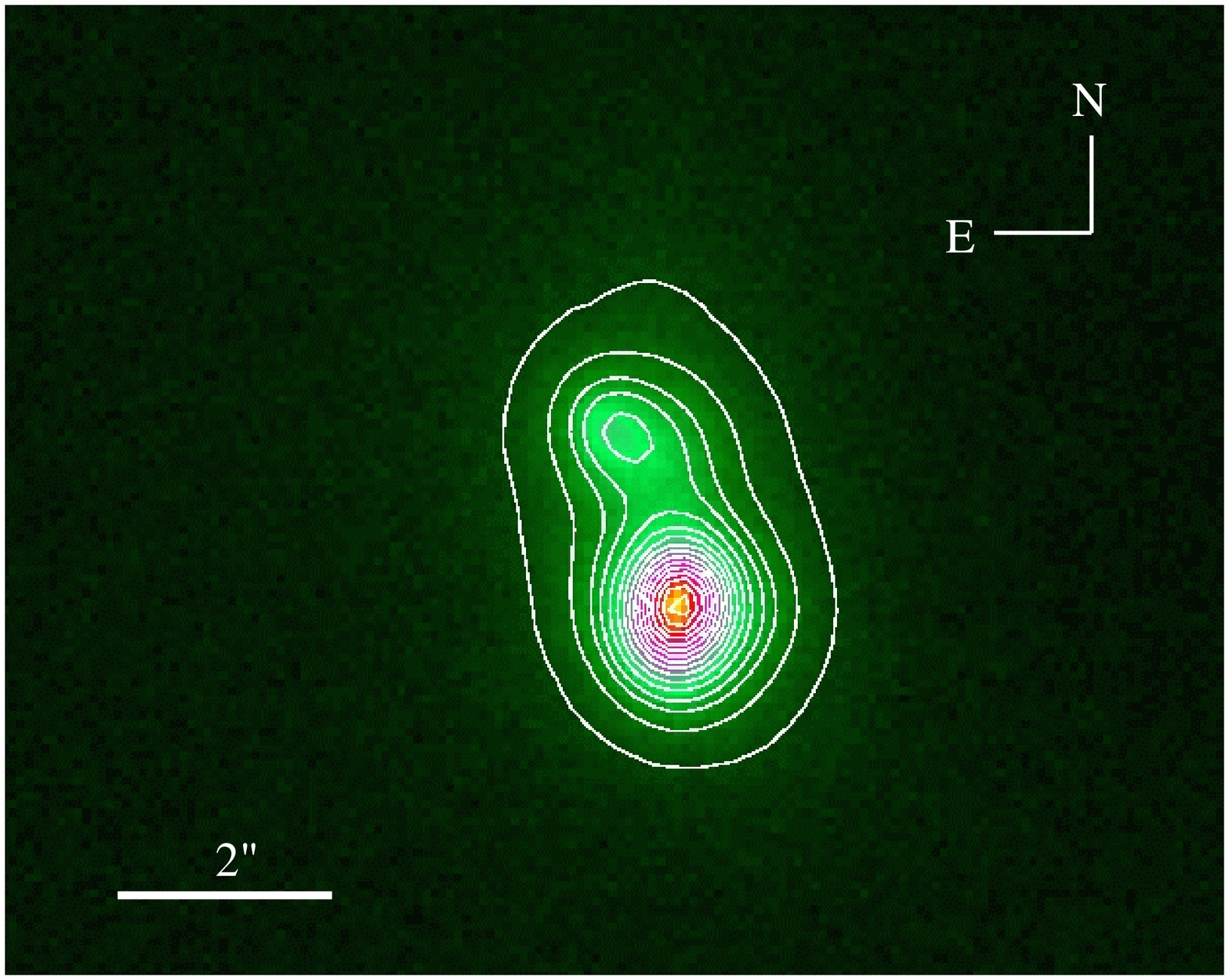}
\figcaption[bogdanovic.fig2a.eps,bogdanovic.fig2b.eps]{Left (a): Narrow band image of the nuclear region of NGC 6240 in the $v=1-0\,S(0)$ transition: the difference between the $H_{2}$ line-plus-continuum intensity (as measured in the $Br\gamma$ filter) and the underlying continuum (as measured in the $H_{2}$ filter). Crosses mark the nuclei as seen in the near-IR continuum image. Right (b): Narrow band image in the continuum emission, with distinct N and S nuclei. The linear scale isocontours are plotted with the step of 5$\%$ in surface brightness where the innermost contour corresponds to the highest value of surface brightness. Image size is 11.34$\times$9.07 arcseconds.\label{narrow_band}}
\end{figure}

	The narrow band images (Figure~\ref{narrow_band}) allow to
	observe the continuum emission and emission from molecular
	lines separately. They show that what is seen as emission from
	the continuum is consistent with the emission coming from the
	active nuclei, while molecular line emission is coming from
	the surrounding gas component.

	 The narrow band image in the 2.12$\,{\rm \mu m}$ rest frame
	 $H_{2}$ $v=1-0\,S(1)$ (hereafter $1-0S\,(1)$ only) transition
	 (Fig~\ref{narrow_band}a) reveals diffuse emission of a gas
	 component. \citet{Tacconi} presented the detailed study of
	 the $H_{2}$ gas dynamics in the nuclear region, traced in the
	 CO millimetar emission. The CO emission maps revealed the
	 existence of a large concentration of molecular gas
	 (2-4$\times 10^{9}\Msun$), in between the two nuclei,
	 confined to a geometrically and optically thick disk, where
	 the dust in the disk efficiently shields the emission from
	 the molecular hydrogen gas. The $H_{2}$ emission seen in our
	 images then probably originates from the parts of nuclear
	 region outside the circumnuclear disk, and from the orders of
	 magnitude less mass of molecular hydrogen than in the disk.

\begin{figure}
\epsscale{1.0}
\plotone{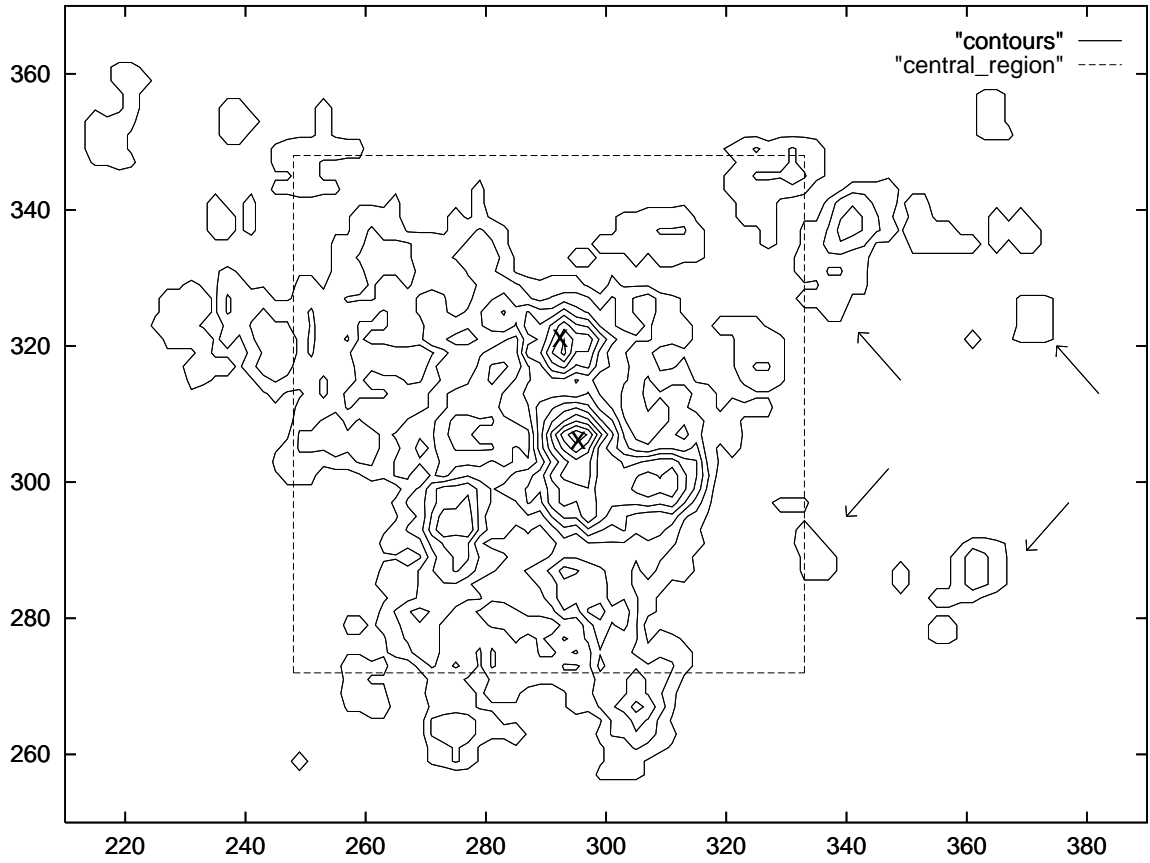}
\figcaption[bogdanovic.fig3.eps]{The X-ray logarithmic isocontour image showing bubbles and filaments of emitting gas on a larger scale (23.7$\times$18.7 arcseconds). A smaller box marks the central region (11.3$\times$9.1 arcseconds) as seen in the near-IR images in Fig.~\ref{narrow_band}. Crosses mark locations of the nuclei as seen in near-IR continuum image. North is up and East is to the left.\label{xray_contour}}
\end{figure}

	 The diffuse emission region extends $400$ to $800\,$pc east
	 and west from the long axis of the system connecting the two
	 nuclei. The strongest $H_{2}$ $1-0\,S(1)$ emission is
	 observed at the position of the S nucleus and between the two
	 nuclei. The peak of the $H_{2}$ emission at the S nucleus
	 appears about ~0.3 arcseconds displaced from the center of
	 the nucleus (as determined from the near-IR continuum image,
	 Fig~\ref{narrow_band}b) towards the north-eastern rim of the
	 S nucleus. This is consistent with the previous detections of
	 the $H_{2}$ $1-0\,S(1)$ emission in the region between the
	 two nuclei \citep{Herbst,vdWerf,Sugai,Tecza} and implies that
	 emission from this gas component is probably not powered by
	 the X-rays coming from the nuclei, but rather by the
	 circumnuclear shocks. A gaseous blob seen in molecular
	 hydrogen emission, located at the $\sim$1kpc distance,
	 south-west to the S nucleus, coincides with the blob emission
	 as seen in the X-ray contour image (Figs~\ref{narrow_band}a
	 and \ref{xray_contour}). The size of the blob is about
	 700$\,$pc. The same region is not seen in the continuum
	 emission which appears associated with the nuclear activity
	 (Fig~\ref{narrow_band}b). This indicates that emission of the
	 gaseous blob may be driven by the local starburst activity,
	 nested inside this region. The N nucleus exhibits low
	 brightness in the $1-0\,S(1)$ molecular transition, while it
	 is distinct on the continuum image. We conclude that the N
	 nucleus is a more quiescent one with a lower rate of
	 starburst and shock waves.

	We compare our near-IR images to the H$\alpha$ images of the
	nuclear region in \citet{Lira} and \citet{Zezas}. We find that
	the general geometries of the extended emission in these two
	images do not resemble each other. Near-IR images show more
	centralized emission on the scale of $\sim$3-4 arcseconds. The
	only example of very faint filamentary structure seen in the
	$1-0\,S(1)$ near-IR image is stretching south-east to the
	nuclear region. The filament is not observed in the continuum
	emission nor the X-ray contour image. Near-IR emission is thus
	spatially more associated with the nuclear activity and the
	circumnuclear starburst than with the larger scale outflows
	and superwinds as further shown in the X-ray contour image
	(Fig~\ref{xray_contour}). The hourglass symmetry of the region
	\citep{Heckman1} and the cone-like outflows \citep{Lira},
	typical of superwinds, and seen in H$\alpha$ images, can not
	be uniquely identified on the $1-0\,S(1)$ image.

 	Both nuclei appear brighter and more distinct in the near-IR
 	$H_{2}$ emission than visible H$\alpha$ emission since the
 	near-IR line is less affected by extinction. The extinction
 	toward the nuclei, estimated from IR observations
 	\citep{Tecza}, is found to be $A_{V}^{N}=5.8$ mag and
 	$A_{V}^{S}=1.6$ mag. The extinction toward the extended
 	emission, estimated from optical observations
 	\citep{Thronson}, is $A_{V}\approx1$ mag.

	From the comparison with the recent Chandra ACIS-S images
	\citep{Komossa1} we notice that the morphology of the near-IR
	continuum image (Fig~\ref{narrow_band}b) corresponds closely
	to that of the hardband (6.0-6.7 keV) X-ray image of the same
	region, showing $\sim$4 arcsecond scale nuclear emission. The
	near-IR 1-0 S(1) image (Fig~\ref{narrow_band}a) corresponds in
	scale and morphology to an ACIS-S intermediate energy band
	image (1.5-2.5 keV), showing more diffuse emission around the
	nuclear region. Finally the morphology of the ACIS-S soft
	energy band image (0.5-1.5 keV) does not correspond to any of
	near-IR images but matches well the H$\alpha$ HST image
	\citep{Lira} in both, scale ($\sim20$ arcsec) and
	morphology. The morphological analogies between X-ray and
	near-IR band suggest that the same mechanisms may be
	responsible for the emission unless the structure is
	significantly under-resolved in one of the bands.

\subsection{X-ray Contours}

	Figure~\ref{xray_contour} is an X-ray logarithmic isocontour
	of the same region observed with Chandra HRC in the energy
	range of 0.08$-$10$\,$keV. The image shows the X-ray emission
	on the scale of $23.7\times 18.7$ arcseconds. The central
	region, marked in the dashed line as a smaller box ($\sim
	11.3\times 9.1$ arcseconds), also shown in
	Figure~\ref{narrow_band}, is a part of a larger scale
	structure which is presented here for clarity.

	The contour plot is made in the logarithmic scale to emphasize
	the faint, extended emission dominated by the nuclear
	emission. The rich gaseous structure seen in the X-ray
	contours matches well the size and morphology of the structure
	seen in the $H\alpha$ emission images \citep{Zezas,Lira}.

	X-ray contours show both nuclei in emission. The nuclei are in
	the north-south orientation with a positional angle of $\sim
	15$ degrees and an angular separation of 1.4 arcseconds. Also
	marked on the X-ray contour image are locations of the nuclei
	as observed in near-IR continuum. The near-IR angular
	separation of nuclei measured is 1.8 arcseconds. The cone-like
	outflows and filaments expanding into the surrounding medium
	are noticeable to the west of the nuclear region (as marked
	with arrows in Figure~\ref{xray_contour}). Visible in the
	X-ray, these are likely to be superwinds capable of heating
	the medium and initiating a new sequence of starburst. The
	emission to the east of the nuclear region is more extended
	but appears to have less filamentary structure.

\subsection{Observed Spectral Features}

\begin{figure}
\epsscale{0.9}
\plotone{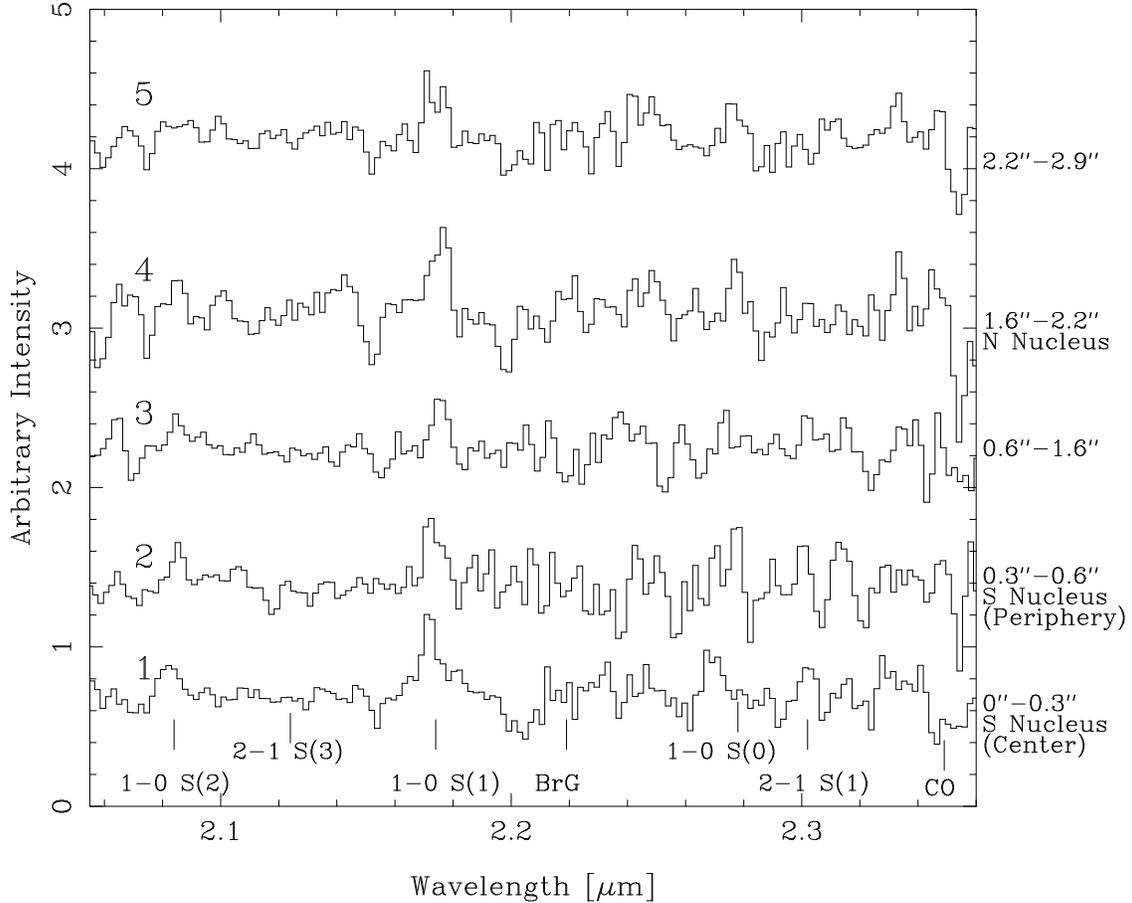}
\figcaption[bogdanovic.fig4.ps]{Observed spectral frames, associated with 5 different parts of nuclear region, as marked on Fig. 1. The width and the position of each region is marked in arcseconds, with zero position defined by the southernmost slit position. The wavelength pixels shown in the spectra are the actual IRCAL pixels.\label{spectra}}
\end{figure}

	 All of our IRCAL frames show a prominent molecular hydrogen
	 $1-0\,S(1)$ line (Fig~\ref{spectra}). This line is formed as
	 a forbidden quadrupole transition at $\lambda=2.12\,{\rm \mu
	 m}$. It can be excited collisionally at temperatures larger
	 than $1000\,$K or radiatively by UV resonant fluorescence.

	 CO(2-0) absorption, noticeable for all slit positions, is
	 characteristic of late spectral type (K and M) giants and
	 supergiants. The typical age of these stars is about $10^{7}$
	 years. This is consistent with a starburst initiated about
	 $10^{7}$ ago, around both nuclei and in the middle region,
	 most likely as a consequence of tidal interaction
	 \citep{Tecza,Ohyama}. Today, these evolved stars contribute a
	 significant amount of radiation to the infrared. Due to the
	 finite spectral resolution we are only able to place an upper
	 limit on the velocity dispersion, $\sigma <750\,{\rm
	 km\,s^{-1}}$, and we do not observe a significant change in
	 the CO(2-0) absorption width towards the N
	 nucleus. \citet{Tecza} measure $\sigma\approx360\,{\rm
	 km\,s^{-1}}$ and $270\,{\rm km\,s^{-1}}$ for N and S nucleus,
	 respectively, with a spectral resolution of R=2000 and a
	 spatial resolution of 1 arcsecond. After correction for
	 inclination of the nuclei, they find dynamical masses of N
	 and S nuclei to be about $8\times 10^{9} \Msun$ and $4\times
	 10^{9} \Msun$, respectively. These values imply that we are
	 witnessing a merger of the nuclei with very massive bulges.

	We also observe $1-0\,S(2)$, $1-0\,S(0)$ and $2-1\,S(1)$
	transitions (Table~\ref{table3}). The relative intensities of
	these lines are in general determined by thermal mechanisms
	and ultraviolet fluorescence. Excitation by ultraviolet
	fluorescence is expected to result in larger populations in
	vibrational levels with $v\ge2$ than for collisional
	excitation in molecular shocks. In the regions where $H_{2}$
	is excited both by collisions at high temperatures
	($T=1000-2000\,K$) and UV sources, lines in the $2{\rm \mu m}$
	region will be dominated by collisional excitation. As
	calculated by \citet{Black1}, the amount of $H_{2}$ necessary
	to produce a certain line intensity at $2{\rm \mu m}$ by
	thermal excitation is an order of magnitude less than that
	required by UV excitation. However, the intensities of lines
	in $v=2$ transitions are sensitive to UV so they are enhanced
	by factors of 3 by UV fluorescence and the excitation effects
	are considerably larger. This implies that detection of the
	radiatively excited transitions is difficult but not
	impossible.

        The emission associated with associative detachment mechanism
        is characteristic for $v=1,2$ vibration levels and high
        rotational levels. The latter allows this mechanism to be
        discriminated from two other competing processes: collisional
        and fluorescent emission. In high density regions, molecules
        are de-excited collisionally which then leads to
        thermalization of the molecular population at low rotational
        levels. In that case associative detachment lines are
        suppressed and it is hard to distinguish it from the other two
        mechanisms. Formation pumping depends strongly on the
        properties of the gas and the $H$, $H^{-}$ reaction cross
        section. These are not well constrained and the contribution
        of this mechanism is hard to estimate \citep{Mouri}.

\subsection{Line Ratios: Comparison of Observations with Model Predictions}

	We calculate line ratios for detected molecular transitions
	and list them in Table~\ref{table4}. The flux units in the
	shown spectra are arbitrary. Consequently the intensities of
	observed emission lines are expressed relative to the
	strongest transition $1-0\,S(1)$, which has an assumed
	intensity of 1.0. Therefore the values for line ratios listed
	in Table 4 are identical with relative line intensities. Line
	ratios for transitions detected with less then 3$\sigma$
	confidence are put as upper limits. This is always determined
	by the detection confidence of the weaker line, since the
	$1-0\,S(1)$ transition is always detected with confidence
	higher than 3$\sigma$.

	Line ratios measured at box locations 1, 2, 3, 4, and 5 (see
	Table~\ref{table1}), are compared against model-predicted line
	ratios, calculated for cases of pure fluorescent excitation
	and pure thermal excitation \citep{Black1} and ``mixture''
	cases: mixture of associative detachment plus fluorescent
	excitation and thermal plus fluorescent excitation for two
	different UV field strengths \citep{Black2}.

	The first model is purely fluorescent (also refered to as
	UV pumping) and calculated line ratios are functions of total
	density $n_{H}$, temperature T, incident UV flux and visual
	extinction $A_{V}$. The model-predicted values do not
	constrain physical parameters for the nuclear region of NGC 6240
	but indicate the observable trend in line ratios. In the case
	of fluorescent excitation as the dominant excitation mechanism,
	line ratios typically have higher values, close to 0.50.  The
	second model is calculated for purely thermal excitation
	at 2000K and has typically lower values of line ratios with
	respect to fluorescence model.
		
 	The third and fourth model indicate combined effects of
 	associative detachment and fluorescent excitation for the gas
 	illuminated by blackbody radiation for blackbody temperatures
 	of 1000K and 5000K. The line ratio predictions in the $v$=1-0
 	band for these two models are close to the values predicted by
 	the thermal model. Unless the $v$=1-0 band ratios are measured
 	with great precision the only sure discriminant is the $v$=2-1
 	band.

	Models five and six predict line ratios from collisional and
	fluorescent excitation of a gas illuminated by radiation from
	a T=1000K and 5000K blackbodies. The rate of radiative
	excitation versus collisional excitation by electrons is the
	same for both models. The values of the line ratios exhibit
	degeneracy with other models, except for the
	$2-1\,S(1)/1-0\,S(1)$ ratio. Model-predicted line ratios for
	$1-0\,S(0)/1-0\,S(1)$ are not available.

	We find that all the $1-0\,S(2)/1-0\,S(1)$ measured line ratios
	are consistent with purely thermal excitation or mixture of
	thermal and fluorescent excitation. Some of the ratio values,
	within the error bars, could be consistent with associative
	detachment+fluorescence mechanism. The purely fluorescent
	model can be eliminated based on this set of ratios, since
	there is only one ratio that marginally supports it.

	The values for the line ratios $1-0\,S(0)/1-0\,S(1)$ and
	$2-1\,S(1)/1-0\,S(1)$, show the same consistency with the
	purely thermal or thermal+fluorescence models, and few of them
	with the associative detachment+fluorescence model. The purely
	fluorescent model is not favored scenario based the values of
	line ratios.

	The Br$\gamma$ emission line, a tracer of young stellar
	population, is not seen in our near-IR spectra. Br$\gamma$ has
	been observed by other authors as a very weak transition with
	the intensity of about 0.07 relative to the intensity of the
	1-0\,S(1) transition \citep{Sugai,Ohyama}. Similarly, we do
	not detect $2-1\,S(2)$ and $2-1\,S(3)$ transitions in our
	spectra (not listed in Table~\ref{table4}) but they were
	detected by \citet{Sugai} and \citet{Ohyama} with relative
	intensities of 0.03 and 0.08, respectively. The fluorescence
	model-predicted ratios for these transitions are $0.28$ for
	$2-1\,S(2)/1-0\,S(1)$ and $0.35$ for
	$2-1\,S(3)/1-0\,S(1)$. The T=2000K thermal model predicts
	$0.03$ and $0.08$ ratios, respectively \citep{Black1}. The
	measured line ratios are therefore closely consistent with the
	predicted values for this thermal model and exclude the
	possibility for significant contribution from the X-ray
	heating, which would result in weaker relative intensities of
	these lines \citep{Sugai}.

	The line ratios are a necessary tool to determine the nature
	of the emission mechanism. In the case of NGC 6240 line ratios
	imply that the whole nuclear region is excited thermally while
	there may be some contribution from the radiative excitation
	mechanism. However the line ratios do not necessarily
	discriminate among the excitation progenitors. For example,
	thermally excited transitions can be powered by heating from
	red giant stars or circumnuclear shocks but also by high
	energy electrons, created by ionization from the central X-ray
	source. On the other hand, fluorescent radiatively powered
	transitions are characteristic of young stellar populations
	only. For $N_{H}\simeq 10^{21-22}\,{\rm cm^{-2}}$ measured
	column density \citep{Komossa1} high energy photons emitted
	from the central source are expected to be mostly thermalized,
	so they do not participate in radiative excitation in the
	near-IR band. By the same argument all radiative transitions
	are likely to come from smaller optical depths. Otherwise,
	they are obscured by gas and dust and thermalized during
	radiative transfer.

\section{Discussion}

	We notice the following similarities in the apparent
	morphology of the nuclear region as seen in different spectral
	bands; near-IR, X-ray observed by \citet{Komossa1} and optical
	observed by \citet{Lira}: a) The near-IR continuum image
	corresponds to the 6.0-6.7 keV band, Chandra X-ray image; b)
	The near-IR 1-0S(1) image corresponds to the 1.5-2.5 keV band
	Chandra image; while c) the Chandra 0.5-1.5 keV image matches
	well the HST H$\alpha$ image. This correlation indicates that
	the same mechanism may be responsible for the emission in
	bands with apparent morphological similarities.

	The near-IR continuum emission and hard band 6.0-6.7 keV X-ray
	emission can be associated with the compact
	emission from the two nuclei. The hard X-ray nuclei ionize the
	surrounding gas and create high energy electrons which further
	participate in collisions with hydrogen atoms and molecules.

	The diffuse 1-0\,S(1) molecular emission, as well as the
	intermediate X-ray energy band emission, are associated with
	the circumnuclear region. The near-IR line ratios indicate
	that the 1-0\,S(1) transition in this region is predominantly
	excited by a thermal mechanism. Circumnuclear shocks are
	thought to be carriers of thermal excitation. The shocked
	$H_{2}$ component probably originates from the volume
	different than the volume occupied by the thick gaseous disk,
	from which stars are forming \citep{Tacconi}. In parts of nuclear
	region, a mixture of thermal and fluorescent excitation may also
	take place, the latter tracing the presence of young stellar
	population capable of UV heating. UV fluorescence is expected
	to originate at smaller optical depths and therefore trace the
	star formation only at a certain physical depth in the nuclear
	region, meaning that the young stellar population could
	potentially be larger than observed from radiatively excited
	transitions.

	Based on the line ratio analysis, certain contributions from
	the associative detachment can not be excluded. Furthermore,
	associative detachment is expected in regions where fast
	shocks pass through the molecular gas. Fast shocks that
	contribute their energy to near-IR, through molecular
	emission, are also likely to contribute to X-ray emission,
	which is observed in Chandra ACIS-S intermediate X-ray energy
	band.

	Optical H$\alpha$ emission is observed on the angular scale of
	20 arcseconds, much larger than $\sim$3-4 arcsecond scale of
	near-IR molecular line emission. Morphologically similar to
	soft X-ray band emission, the H$\alpha$ emission is likely to
	be associated with large scale outflows and superwinds.

	The morphological comparison of the nuclear region as seen in
	different energy bands is valid under the assumption that
	dominant features in the nuclear region are resolved in all of
	the bands. This can be questioned if the longer imaging
	integration would reveal significantly different scale and
	features in the nuclear region, from those in images we refer
	to.

	In the whole nuclear region we also detect the presence of an
	aged starburst population, initiated $\sim10^7$yr ago probably
	as a consequence of tidal interaction of two nuclei
	\citep{Lester1}. These are mostly found to be late type
	supergiants that contribute all of the CO absorption and a
	fraction of thermally induced emission to $2\,{\rm \mu m}$
	band transitions.

\section{Conclusions}

	In this work we base our conclusions on high spatial
	resolution near-IR imaging and spectroscopy and available
	X-ray imaging. NGC 6240 is revealed as a luminous galaxy with
	multiple components in emission. We find that near-IR
	luminosity emitted from the nuclear region of NGC 6240 is
	spatially associated with the central kiloparsec (3-4
	arcsecond scale). We observe both nuclei in emission with the
	S nucleus being the more prominent. We also observe the
	diffuse emission from the gas component distributed around and
	between the nuclei.

	The values of measured line ratios in near-IR are consistent
	with {\it thermal excitation} as the dominant excitation
	mechanism. Thermal emission can be driven by collisional
	excitation in small scale circumnuclear shocks and also
	contributed by late type, giant and supergiant stars. A
	smaller fraction of energy can be contributed radiatively,
	through {\it UV fluorescence} and comes from young stellar
	population. The {\it associative detachment} emission
	mechanism can not be excluded based on our data. This
	mechanism is likely to contribute to near-IR emission, since
	it is typically triggered in molecular regions with presence
	of fast shocks. The {\it X-ray radiation} emitted by the two
	active nuclei under 10 keV is scattered and thermalized. Its
	contribution to near-IR band is the most noticeable in
	continuum emission, associated with the N and S
	nuclei. Near-IR data, compared to X-ray and optical data
	available, indicate that thermal excitation is the dominant
	mechanism but not the only one.

\acknowledgements
 We are indebted to J.R. Graham for allowing us to use IR camera and
 S.A. Severson, J.P. Lloyd and D.T. Gavel for the observing. We are
 very grateful to W.N. Brandt for his help with the X-ray image. This
 work is supported in part by Penn State Eberly College of Science,
 National Science Foundation with grant AST-0138235, NASA with grants
 NAG5-12115 and NAG5-11427, and under the auspices of the
 U.S. Department of Energy, National Nuclear Security Administration
 by the University of California, Lawrence Livermore National
 Laboratory under contract No. W-7405-Eng-48. This work was supported
 in part by the National Science Foundation Science and Technology
 Center for Adaptive Optics, managed by the University of California
 at Santa Cruz under cooperative agreement No. AST-9876783. We
 acknowledge anonymous referee for useful suggestions.


\clearpage

\clearpage

\clearpage

\clearpage

\begin{table}
\begin{center}
\begin{tabular}{cccccc} 
\tableline\tableline
Spectrum & Obs. Date & Exposure & Area & Std. Star & S/N \\ (1) & (2)
& (3) & (4) & (5) & (6) \\
\tableline
 1 & 2000 Jun 25 & 600 & 0.3$\times$5 & HD203856 & 40\\ 
 2 & 2000 Jun 25 & 1200 & 0.3$\times$5 & HD203856 & 15\\ 
 3 & 2000 Jun 25 & 2100 & 1.0$\times$5 & HD203856 & 15\\ 
 4 & 2000 Jun 25 & 1500  & 0.6$\times$5 & HD203856 & $\le$10\\ 
 5 & 2000 Jun 25 & 1500  & 0.7$\times$5 & HD203856 & 10\\ 
\tableline
\end{tabular}
\caption{Columns: (1) Spectrum as marked on Fig. 1; (2) Observing date; (3) Exposure time in seconds; (4) Area on the sky in square arcseconds; (5) Standard star used in spectral reduction; (6) Signal to noise ratio. \label{table1}}
\end{center}
\end{table}

\clearpage

\begin{table}
\begin{center}
\begin{tabular}{cccccccc}
 \tableline\tableline
 Image & Obs. Date & Exposure & Size & Filter & $\lambda_{c}$ & FWHM & S/N \\ 
(1) & (2) & (3) & (4) & (5) & (6) & (7) & (8) \\
\tableline
 Fig.1 & 2000 Jun 25 & 60  & 11.34$\times$9.07 & Ks & 2.150 & 0.320 & 110 , 38 \\
 Fig.2a  & 2001 Aug 6 & 2633  & 11.34$\times$9.07 & $Br\gamma-H_{2}$ & 2.167 & 0.020 & 6400 , 550 \\  
 Fig.2b & 2001 Aug 6 & 1772 & 11.34$\times$9.07 & $H_{2}$ & 2.125 & 0.020 & 5300 , 560\\ 
\tableline 
\end{tabular}
\caption{Columns: (1) Image name; (2) Observing date; (3) Exposure time in seconds; (4) Size of image in figure; (5) Filter; (6) Central wavelength in microns; (7) Full width half maximum; (8) Signal to noise ratio for compact and diffuse emission. \label{table2}}
\end{center}
\end{table}

\clearpage

\begin{table}
\begin{center}
\begin{tabular}{crcc} 
\tableline\tableline
Emitter & Transition & Rest  (${\rm \mu m}$) & Obs. (${\rm \mu m}$)\\ 
\tableline 
$H_{2}$ & v=1-0S(2) & 2.0338 & 2.0836\\  
$H_{2}$ & v=2-1S(3) & 2.0735 & 2.1243\\ 
$H_{2}$ & v=1-0S(1) & 2.1218 & 2.1738\\ 
$H_{2}$ & v=2-1S(2) & 2.1542 & 2.2070\\ 
Br$\gamma$ & ... & 2.1661 & 2.2192\\ 
$H_{2}$ & v=1-0S(0) & 2.2233 & 2.2778\\ 
$H_{2}$ & v=2-1S(1) & 2.2477 & 2.3028\\ 
\tableline
\end{tabular}
\caption{Emission line transitions in rest and observed frame.\label{table3}}
\end{center}
\end{table}

\clearpage

\clearpage

\begin{table}
\begin{center}
\begin{tabular}{cccc}
\tableline\tableline
Observed/Modeled & 1-0S(2)/1-0S(1) & 1-0S(0)/1-0S(1) & 2-1S(1)/1-0S(1) \\ 
\tableline
1\tablenotemark{a} & 0.36$\pm$0.12 & ... & $<$0.37\\ 
2\tablenotemark{a} & 0.38$\pm$0.16 & $<$0.38 & $<$0.34\\ 
3\tablenotemark{a} & 0.24$\pm$0.10 & ... & $<$0.42 \\ 
4\tablenotemark{a} & $<$0.32 & $<$0.36 & ... \\ 
5\tablenotemark{a} & $<$0.40 & 0.45$\pm$0.23 & ... \\ 
&&&\\
Fluorescent\tablenotemark{b} & 0.50 & 0.45 & 0.56\\
Thermal (2000K)\tablenotemark{b}& 0.37 & 0.21 & 0.082\\
Associ Detach. + Fluorescent (1000K)\tablenotemark{c} & 0.41 & 0.43 & 0.55\\
Associ. Detach. + Fluorescent (5000K)\tablenotemark{c} & 0.42 & 0.32 & 0.53\\
Thermal + Fluorescent (1000K)\tablenotemark{d}& 0.41 & ... & 0.54\\
Thermal + Fluorescent (5000K)\tablenotemark{d}& 0.37& ... & 0.19\\
\tableline 
\end{tabular} 
\tablenotetext{a}{Observed line ratios. See text.}
\tablenotetext{b}{Model 14 of Black \& van Dishoeck 1987.}
\tablenotetext{c}{Combined effects of associative detachment and UV pumping, Black \& van Dishoeck 1981.}
\tablenotetext{d}{Combined effects of thermal and fluorescent excitation, Black \& van Dishoeck 1981.}
\caption{Observed and model-predicted line ratios. Line ratios are 
calculated for lines detected with more than 3$\sigma$ confidence. The
errors reported are 1$\sigma$. The upper limits are calculated for
transitions detected with $> 1.5\,\sigma$ confidence. \label{table4}}
\end{center} 
\end{table} 
\clearpage 

\end{document}